\pdfoutput=1
\documentclass{article}
\usepackage[preprint]{spconf}
\usepackage{amsmath,graphicx}

\copyrightnotice{\copyright\ IEEE 2021}
\toappear{To appear in {\it Proc.\ ICASSP 2021, June 6-11, 2021}}

\usepackage{enumitem}

\usepackage{mathtools}

\usepackage{url}

\usepackage{balance}

\usepackage{siunitx}
\usepackage{cleveref}
\Crefname{section}{Sec.}{Secs}  

\usepackage[acronym]{glossaries}
\usepackage{balance}

\usepackage{tabularx, etoolbox, booktabs}
\usepackage{multirow} 
\newcolumntype{H}{>{\setbox0=\hbox\bgroup}c<{\egroup}@{}}  

\usepackage{subcaption}

\usepackage{algorithm}
\usepackage{algpseudocode}

\algtext*{EndWhile}  
\algtext*{EndIf}  
\algtext*{EndFor}

\usepackage{pgfplots} 
\newlength\fheight 
\newlength\fwidth 
\usepackage{tikz}
\pgfplotsset{compat=1.9}

\usetikzlibrary{external}

\usetikzlibrary{arrows}
\usetikzlibrary{patterns}
\usetikzlibrary{backgrounds}
\usetikzlibrary{fit}
\usetikzlibrary{positioning}
\usetikzlibrary{shapes.geometric}
\usetikzlibrary{calc}
\usetikzlibrary{shapes.multipart}
\usetikzlibrary{shapes.misc}

\tikzset{>=stealth}

\tikzstyle{block}=[
    draw,
    text depth=0pt,
    thick, 
    rectangle,
    text centered,
    minimum height=3.4ex,
    fill=black!6,
    ] 

\tikzstyle{branch}=[{circle,inner sep=0pt,minimum size=0.3em,fill=black}]
\tikzstyle{box}=[rectangle, rounded corners, draw=black, line width=1pt, text width=2cm]

\tikzstyle{arrow}=[{}-{>}, thick]
\tikzstyle{reverse arrow}=[{<}-{}, thick]

\tikzset{
	do path picture/.style={%
		path picture={%
			\pgfpointdiff{\pgfpointanchor{path picture bounding box}{south west}}%
			{\pgfpointanchor{path picture bounding box}{north east}}%
			\pgfgetlastxy\x\y%
			\tikzset{x=\x/2,y=\y/2}%
			#1
		}
	},
	sin wave/.style={do path picture={    
			\draw [line cap=round] (-3/4,0)
			sin (-3/8,1/2) cos (0,0) sin (3/8,-1/2) cos (3/4,0);
	}},
	cross/.style={do path picture={    
			\draw [line cap=round] (-2/5,-2/5) -- (2/5,2/5) (-2/5,2/5) -- (2/5,-2/5);
	}},
	plus/.style={draw, circle, do path picture={    
			\draw [line cap=round] (-3/5,0) -- (3/5,0) (0,-3/5) -- (0,3/5);
	}},
	mic/.style={inner sep=0pt, do path picture={
			\draw (0,0) circle (0.9);
			\draw [line cap=round] (-0.9, -0.9) -- (-0.9, 0.9);
	}},
	mux/.style={trapezium, draw}
}

\usepackage{amsmath}
\usepackage{amssymb}
\usepackage{bbm}
\usepackage{physics}
\usepackage{siunitx}
\usepackage{trfsigns}

\DeclareMathOperator*{\argmin}{argmin}

\newcommand{\vect}[1]{\ensuremath{\boldsymbol{\mathbf{#1}}}}

\renewcommand{\H}{^\mathrm{H}}

\def\x{{\mathbf x}}

\newacronym{ASR}{ASR}{Automatic Speech Recognition}
\newacronym{ATF}{ATF}{acoustic transfer function}
\newacronym{BF}{BF}{beamforming}
\newacronym{BLSTM}{BLSTM}{Bidirectional Long Short-Term Memory}
\newacronym{CNN}{CNN}{Convolutional Neural Network}
\newacronym{CTF}{CTF}{convolutive transfer function}
\newacronym{CISDR}{CI-SDR}{Convolutive transfer function Invariant Signal-to-Distortion Ratio}
\newacronym{EM}{EM}{Expectation Maximization}
\newacronym{GSC}{GSC}{Generalized Sidelobe Canceller}
\newacronym{GMM}{GMM}{Gaussian Mixture Model}
\newacronym{GSS}{GSS}{Guided Source Separation}
\newacronym{ICA}{ICA}{Independent Component Analysis}
\newacronym{IVA}{IVA}{Independent Vector Analysis}
\newacronym{LF-MMI}{LF-MMI}{Lattice-Free Maximum Mutual Information}
\newacronym{MIMO}{MIMO}{Multiple Input Multiple Output}
\newacronym{ML}{ML}{maximum likelihood}
\newacronym{MLDR}{MLDR}{Maximum Likelihood Distortionless Response}
\newacronym{MM}{MM}{Mixture Model}
\newacronym{MSE}{MSE}{Mean Square Error}
\newacronym{MPDR}{MPDR}{Minimum-Power Distortionless Response}
\newacronym{MVDR}{MVDR}{Minimum Variance Distortionless Response}
\newacronym{NN}{NN}{neural network}
\newacronym{PSD}{PSD}{Power Spectral Density}
\newacronym{PIT}{PIT}{permutation invariant training}
\newacronym{PESQ}{PESQ}{Perceptual Evaluation of Speech Quality}
\newacronym{RIR}{RIR}{room impulse response}
\newacronym{RTF}{RTF}{relative transfer function}
\newacronym{TDNN}{TDNN}{Time Delay Neural Network}
\newacronym{TDNN-F}{TDNN-F}{factorized TDNN}
\newacronym{STFT}{STFT}{Short Time Fourier Transform}
\newacronym{iSTFT}{iSTFT}{inverse Short Time Fourier Transform}
\newacronym{SDR}{SDR}{Signal-to-Distortion Ratio}
\newacronym{SNR}{SNR}{Signal-to-Noise Ratio}
\newacronym{SISDR}{SI-SDR}{Scale Invariant Signal-to-Distortion Ratio}
\newacronym{STOI}{STOI}{Short Time Objective Intelligibility}
\newacronym{WER}{WER}{Word Error Rate}
\newacronym{wMPDR}{wMPDR}{weighted Minimum-Power Distortionless Response}
\newacronym{WPE}{WPE}{Weighted Prediction Error}
\newacronym{WPD}{WPD}{weighted power minimization distortionless response}

\title{
    Convolutive Transfer Function Invariant SDR training criteria for Multi-Channel Reverberant Speech Separation
}
\name{\begin{tabular}{c}%
Christoph Boeddeker$^{1}$ %
\qquad %
Wangyou Zhang$^{2}$ %
\qquad %
Tomohiro Nakatani$^{3}$ %
\\%
Keisuke Kinoshita$^{3}$ \qquad %
Tsubasa Ochiai$^{3}$ %
\qquad %
Marc Delcroix$^{3}$ %
\qquad %
Naoyuki Kamo$^{3}$ %
\\%
Yanmin Qian$^{2}$ %
\qquad %
\qquad %
Reinhold Haeb-Umbach$^{1}$ %
\end{tabular}}
\address{%
$^{1}$ Paderborn University, Department of Communications Engineering, Paderborn, Germany \\
$^{2}$ SpeechLab, Department of Computer Science and Engineering, Shanghai Jiao Tong University, China \\
$^{3}$ NTT Corporation, Japan \\
}
\makeatletter
\renewcommand\section{\@startsection {section}{1}{\z@}%
                                   {-2.5ex \@plus -1ex \@minus -.2ex}%
                                   {1.75ex \@plus.2ex}%
                                   {\normalfont\Large\bfseries}}
\renewcommand\subsection{\@startsection{subsection}{2}{\z@}%
                                     {-1.5ex\@plus -1ex \@minus -.2ex}%
                                     {1.ex \@plus .2ex}%
                                     {\normalfont\large\bfseries}}
\makeatother

\begin{document}
\ninept
\maketitle
\begin{abstract}
Time-domain training criteria have proven to be very effective for the separation of single-channel non-reverberant speech mixtures.
Likewise, mask-based beamforming has shown impressive performance in multi-channel reverberant speech enhancement and source separation.
Here, we propose to combine neural network supported multi-channel source separation with a time-domain training objective function.
For the objective we propose to use a convolutive transfer function invariant Signal-to-Distortion Ratio (CI-SDR) based loss.
While this is a well-known evaluation metric (BSS Eval), it has not been used as a training objective before.
To show the effectiveness, we demonstrate the performance on LibriSpeech based reverberant mixtures.
On this task, the proposed system approaches the error rate obtained on single-source non-reverberant input, i.e., LibriSpeech test\_clean, with a difference of only 1.2 percentage points, thus outperforming 
a conventional permutation invariant training based system and alternative objectives like Scale Invariant Signal-to-Distortion Ratio by a large margin.
\end{abstract}
\begin{keywords}
Multi-channel source separation, acoustic beamforming, complex backpropagation, Signal-to-Distortion Ratio
\end{keywords}
\section{Introduction}
\label{sec:intro}

Blind speech separation aims at extracting the individual speech signals
present in a mixture.
It is considered an important signal enhancement step both in human-to-human communication and for a downstream \gls{ASR} system. Having been a topic of extensive research for many years, many solutions have been proposed,
such as the \gls{ICA} \cite{Comon1994ICA}, \gls{IVA} \cite{Kim2006IVA}, non-negative matrix factorization \cite{Lee1999NNMF}, spatial mixture model based techniques \cite{Araki2006SMM}, and deep neural network based separation methods, which are  the focus of this contribution.

In a popular variant, the purpose of the \gls{NN} is to estimate time-frequency masks, or, stated differently, a speaker presence probability for each speaker and each time-frequency bin,
thus taking advantage of the sparsity and W-disjoint orthogonality \cite{Yilmaz2004WDO} of speech signals in this domain.
The actual source extraction can be carried out either by applying the mask to a channel of the input signal or by acoustic beamforming.
The latter requires multi-channel input, but is known to lead to perceptually more pleasing results exhibiting less artifacts \cite{Pertila2015distant}.
Furthermore, it takes advantage of spatial information, which can be gleaned from a microphone array, and, thus, usually leads to better word error rates of a downstream \gls{ASR} task \cite{Drude2019integration}.

While earlier publications, such as \gls{PIT} \cite{Kolbaek2017uPIT}, deep clustering \cite{Hershey2016deepClustering} and variants thereof \cite{Wang2018},
employed neural network training criteria that were defined in the \gls{STFT} domain, more recent publications  suggest that loss functions defined in the time-domain, such as the (scale invariant) \gls{SDR}, generally achieve superior separation performance \cite{Luo2018tasnet, Kolbaek2020loss}.
In fact, the investigation in \cite{Heitkaemper2020demystifying} showed that the advantage of time-domain loss functions is maintained even if the mask estimation is actually carried out in the frequency domain.
However, the combination of time-domain \gls{NN} training criteria and source extraction by beamforming at training time is widely unexplored, and will be the focus of this work. 

In this contribution, we consider source separation as a front-end of a downstream \gls{ASR} task.
One might therefore argue that the front-end should best be trained using an ASR-related criterion, as the latter is closer to the ultimate goal of minimal \gls{WER}.
This has been attempted in \cite{Chang2019mimo} where a sequence-to-sequence neural \gls{ASR} system was extended to deal with multi-channel multi-source input. 
While the results on anechoic speech were promising, their performance on reverberant speech was not yet competitive.
For single-source acoustic beamforming the Beamnet architecture \cite{Heymann2017beamnet} has been proposed, which utilizes an \gls{ASR}-related training objective, and the  gradient was backpropagated through the beamformer to the neural mask estimator.
It was, however, observed in \cite{Heymann2018MBBFSmartHome} that joint training of  the \gls{ASR} back-end and the enhancement front-end may have some logistic advantages (no need for parallel clean and distorted training data for the training of the enhancement stage), but that they may not lead to the overall best \gls{WER} performance. We therefore opted to stick to a signal related training objective for the source separation training in this work, which is actually much simpler to realize, and leave a tighter coupling with the back-end to future work.

In this contribution, we focus on reverberant scenarios.
A time-domain loss can be extremely sensitive to changes in the input waveform, e.g., caused by the different reflection pattern when selecting another channel of the microphone array as input to the mask estimation network \cite{Drude2019sms}.
Although inaudible, those changes can have a drastic impact on the loss, and unfavorably influence the performance.
To overcome this problem, we propose a training objective that is invariant to such errors, i.e., \acrfull{CISDR} loss, inspired by the
BSS Eval SDR \cite{Vincent2006BSSEval} measure.
We give experimental evidence that it is superior to the well-known  \gls{SISDR} if combined with an \gls{MVDR}-based beamformer.
On a reverberant source separation task compiled from LibriSpeech data \cite{Panayotov2015librispeech}, which was developed during the JHU  JSALT 2020 workshop, the proposed system achieves a word error rate which is only \num{1.2} percentage points higher than on the non-reverberant oracle source signals.
The \acrshort{CISDR} PyTorch code is publicly available on GitHub\footnote{\url{https://github.com/fgnt/ci_sdr}}.


The paper is organized as follows. In the next section we introduce the signal model underlying our investigations. In Section~\ref{sec:system} the system architecture is introduced. 
To learn the neural mask estimator with a time-domain training criterion, the loss has to be backpropagated through the beamformer coefficient computation. 
We use a beamformer implementation that requires computing the dominant eigenvector to estimate the steering vectors.
We propose using the power estimation to simplify such a computation during training.
Section~\ref{sec:training_objectives} discusses the training objectives, which are experimentally evaluated in Section~\ref{sec:experiments}. The paper closes with some conclusions drawn in Section~\ref{sec:conclusions}.

\newcommand{\y}{\vect{y}}
\renewcommand{\x}{\vect{x}}
\renewcommand{\a}{\vect{a}}
\newcommand{\n}{\vect{n}}
\newcommand{\w}{\vect{w}}
\newcommand{\R}{\vect{R}}
\renewcommand{\v}{\vect{v}}
\renewcommand{\d}{\vect{d}}
\renewcommand{\r}{\vect{r}}

\section{Signal model}

Assuming $I$ concurrent speakers and an array of $M$ microphones, the vector of signals at the microphones, $\y_{\ell}= [y_{\ell, 1}, \ldots , y_{\ell, M}]^\top$, at sampling time $\ell$ can be written as follows
\begin{align}
    \y_\ell
    &= \sum_{i=1}^I \x_{\ell, i} + \n_\ell ,
\end{align}
where
\newcommand{\RIRLength}{L_{\mathrm{\tau}}}
\begin{align}
    \x_{\ell, i}
    = \d_{\ell, i} + \r_{\ell, i}
    = \sum_{\tau = 0}^{\mathclap{\RIRLength^{\mathrm{early}}-1}} \a_{\tau, i} s_{\ell - \tau, i} + \sum_{\mathclap{\tau = \RIRLength^{\mathrm{early}}}}^{\mathclap{\RIRLength-1}} \a_{\tau, i} s_{\ell - \tau, i}.
\end{align}
Here, $\x_{\ell, i}$ is the image of the $i$th source at the microphones and $\n_\ell$ the noise vector. Further,
$\a_{\tau, i}$ is the \gls{RIR} vector from the $i$th source to the microphone array at time lag $\tau$. It can be decomposed in an early and a late part, resulting in an ``early'' part of the image at the microphones, $\d_{\ell, i}$, and a late contribution $\r_{\ell, i}$, where the former contains the direct signal and early reflections (typically up to \SI{50}{ms} of the \gls{RIR}) and the latter the late reverberation.

In the \gls{STFT} domain this model can be approximated as follows
\begin{align}
    \y_{t, f}
    &= \sum_{i=1}^I \d_{t, f, i} + \sum_{i=1}^I \r_{t, f, i} +\n_{t, f} ,
\end{align}
where $t$ and $f$ denote the time frame index and  the frequency bin index, respectively.
The early and late arriving speech are given by 
\begin{align}
\label{eq:def_direct_f}
\d_{t, f, i}
    &= \sum_{\lambda=0}^{\Delta - 1}
     \a_{\lambda, f, i} s_{t-\lambda, f, i}
    \approx \v_{f,i}s_{t,f,i} = \tilde{\v}_{f,i,r} d_{t,f,i,r} ,\\
    \r_{t, f, i} &= \sum_{\lambda=\Delta}^{L_\lambda - 1}
    \a_{\lambda, f, i} s_{t-\lambda, f, i},
\end{align}
where $\Delta$ is set to correspond to approximately \SI{50}{ms}.

Our goal is to extract the early part of the source image $d_{t, f, i,r}$ at a reference microphone $r$ for each source $i$. By doing so, we not only aim at removing the competing speakers $i^\prime \ne i$ and noise, but also wish to remove the late reverberation, which is known to significantly degrade recognition performance. For acoustic beamforming the multiplicative transfer function approximation is applied in eq.~\eqref{eq:def_direct_f}, which also introduces the \gls{RTF} vector $\tilde{\v}_{f,i,r} = \v_{f,i}/v_{f,i,r}$, where $v_{f,i,r}$ is the $r$th component of the steering
vector $\v_{f,i}$. 


\section{Enhancement System Architecture}
\label{sec:system}

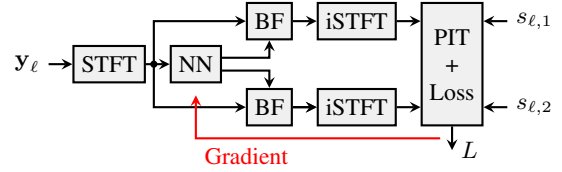
\begin{figure}
    \centering
    \begin{tikzpicture}
        \node[block] (NN) {\acrshort{NN}};
        \node[block, anchor=east] (STFT) at ($(NN.west) + (-1em, 0)$) {\acrshort{STFT}};
        \node[block, anchor=south west] (bf1) at ($(NN.east) + (1em, 1em)$) {BF};
        \node[block, anchor=north west] (bf2) at ($(NN.east) + (1em, -1em)$) {BF};
        \node[block, anchor=west] (iSTFT1) at ($(bf1.east) + (1em, 0)$) {\acrshort{iSTFT}};
        \node[block, anchor=west] (iSTFT2) at ($(bf2.east) + (1em, 0)$) {\acrshort{iSTFT}};
        
        \path let
          \p1 = (iSTFT1.north),
          \p2 = (iSTFT2.south),
          \n1 = {\y1-\y2-2\pgflinewidth} 
          in
        node[block, anchor=west, minimum height=\n1, align=center] (loss) at ($(iSTFT2.east)!0.5!(iSTFT1.east) + (1em, 0)$) {\acrshort{PIT}\\+\\Loss};
        
        \draw[arrow] (STFT) -- (NN);
        \draw[arrow] ($(NN.east)!1/3!(NN.north east)$) -| (bf1);
        \draw[arrow] ($(NN.east)!1/3!(NN.south east)$) -| (bf2);
        \draw[arrow] (bf1) -- (iSTFT1);
        \draw[arrow] (bf2) -- (iSTFT2);
        \draw[arrow] (iSTFT1) -- (loss.west |- iSTFT1);
        \draw[arrow] (iSTFT2) -- (loss.west |- iSTFT2);
        
        \draw[arrow] (bf2) -- (iSTFT2);
        
        \draw[reverse arrow] (STFT.west) -- +(-1em, 0) node[left] {$\y_{\ell}$};
        \draw[arrow] ($(STFT.east)!1/3!(NN.west)$) node[branch]{} |-  (bf1);
        \draw[arrow] ($(STFT.east)!1/3!(NN.west)$) |-  (bf2);
        
        \draw[reverse arrow] (loss.east|-iSTFT1) -- +(1em, 0) node[right] {$s_{\ell, 1}$};
        \draw[reverse arrow] (loss.east|-iSTFT2) -- +(1em, 0) node[right] {$s_{\ell, 2}$};

        \draw[arrow] (loss.south) --  +(0, -1em) node[right]{$L$};
        \draw[arrow, red, shorten >=0.5em,shorten <=0.5em] ($(loss.south) + (0, -0.5em)$) -| node[below right]{Gradient} (NN);

    \end{tikzpicture}

    \caption{System overview. The gradient is backpropagated from the time-domain loss through the inverse \gls{STFT} and beamforming to the \gls{NN} parameters.}
    \label{fig:block_diagram}
\end{figure}

The block diagram of the enhancement system is depicted in Fig.~\ref{fig:block_diagram}. The multi-channel input $\y_\ell$ is transformed to the \gls{STFT} domain, where $I$ beamformers (BF) are used to extract the source signals from the mixture. The masks for the beamformer coefficient computation are estimated by a \gls{NN}, whose input is one of the microphone channels. The loss function for the network training is computed in the time-domain on the beamformed signals and then backpropagated to the NN. 
Note that a source permutation problem occurs in the loss computation because the network output order may not match the  order of the target signals.
To address this, we use the \gls{PIT} loss \cite{Kolbaek2017uPIT}:
We compute the loss for each permutation of the network output and search for its minimum.
Then, we take this minimal loss to compute the gradients for backpropagation.

\subsection{Mask estimation}
 The input to the \gls{NN} is  the log magnitude of the \gls{STFT}  (with size $1024$ and shift $256$ at \SI{16}{kHz}) of the microphone signal at a reference channel $r$: $\log(1 + |y_{f, t, r}|)$.
 The constant is added to avoid $\log(0)$.

The \gls{NN} consists of 3 \gls{BLSTM} layers with $600$ hidden units in each direction, followed by two feed forward layers.
The first feed forward layer keeps the feature size and the second expands it to $512 \cdot 3 \cdot I$.
That is, the network outputs three masks per speaker: $m_{\d, t, f, i}$, $m_{\n, t, f, i}$, and $m_{\tilde{\n}, t, f, i}$.
They are used for the estimation of 
\begin{itemize}[leftmargin=*]
    \setlength\itemsep{0em}
    \item The spatial covariance matrix of the target speaker $\R_{\d, f, i}$,
    \item The spatial covariance matrix of the distortions (competing speakers plus noise) $\R_{\n, f, i}$ of target speaker $i$, to be used in the MVDR beamformer computation (see Section~\ref{sec:mvdr}), and
    \item The spatial covariance matrix of the distortions $\R_{\tilde{\n}, f, i}$ to be used in the dominant eigenvector estimation (see Section~\ref{sec:rtf}).
\end{itemize}
The covariance matrices are computed as follows:
\begin{align}
    \R_{\boldsymbol{\nu}, f, i} &= \frac{1}{T} \sum_{t} (\varepsilon + m_{\boldsymbol{\nu}, t, f, i}) \y_{t, f}\y_{t, f}\H
    \label{eq:covest} ,
\end{align}
where $\boldsymbol{\nu} \in \{\d, \n, \tilde{\n}\}$, and where $\varepsilon = 0.01$ is a small value to improve the training convergence.

Clearly, one could use $\R_{\n, f, i} = \R_{\tilde{\n}, f, i}$, but we decided to make this distinction because it showed slightly better performance.


\subsection{MVDR beamformer}
\label{sec:mvdr}
For source extraction we employ the well-known \gls{MVDR} beamformer \cite{Ban1988beamforming}:
\begin{align}
    \w_{f, i, r} &= \frac{\R_{\n, f, i}^{-1}\tilde{\v}_{f, i, r}}{\tilde{\v}_{f, i, r}\H\R_{\n, f, i}^{-1}\tilde{\v}_{f, i, r}} , \\
    \hat{d}_{t, f, i, r} &= \w_{f, i, r}\H \y_{t, f} ,
\end{align}
where $\w_{f, i, r}$ is a vector containing the beamformer coefficients and $\hat{d}_{t, f, i, r}$ is the estimate of the desired signal at the reference channel $r$. 

\subsection{Relative transfer function estimation}
\label{sec:rtf}
For the \gls{RTF} estimation we follow \cite{Ito2017MaxEigATF}:
\begin{align}
    \label{eq:rtfest}
    \v_{f, i} &= \R_{\tilde{\n}, f, i} \operatorname{MaxEig}\left\{ \R_{\tilde{\n}, f, i}^{-1} \R_{\d, f, i} \right\} ,\\
    \tilde{\v}_{f, i, r} &= \v_{f, i} / v_{f, i, r},
\end{align}
where $\operatorname{MaxEig}\{\cdot\}$ extracts the eigenvector corresponding to the largest eigenvalue.
However, this approach has a theoretical and practical drawback: First, the gradient is complicated \cite{Boeddeker2017NNBF}.
For this work we used PyTorch and at the time of writing, the support for complex numbers was not yet finished.
Therefore, each complex operation was mapped to real operations\footnote{\url{https://github.com/kamo-naoyuki/pytorch_complex}}.
However, for the eigenvalue decomposition we couldn't find such a mapping that worked. See \cite{Boeddeker2017NNBF} for a discussion of some of the issues.

As an alternative we employed the power iteration algorithm to obtain the eigenvector corresponding to the dominant eigenvalue \cite{Mises1929PowerIteration}.
The algorithm to estimate the \gls{RTF} is then:
\begin{algorithmic}[1]
    \State $\vect{\Phi}_{f, i} = \R_{\tilde{\n}, f, i}^{-1} \R_{\d, f, i}$
	\State ${\v}_{f, i} \gets \vect{u}_r$ \quad \# $\vect{u}_r$ \dots one hot vector, i.e., $r$th component is $1$
	\For{$\eta \in \left(1, \dots, \eta_{\text{max}}\right)$}
		\State ${\v}_{f, i} \gets \vect{\Phi}_{f, i} {\v}_{f, i}$
	\EndFor
	\State ${\v}_{f, i} \gets \R_{\tilde{\n}, f, i} {\v}_{f, i}$ 
	\State $\tilde{\v}_{f, i, r} = {\v}_{f, i} / v_{f, i, r}$ 
\end{algorithmic}
where line 2 to line 4 is the power iteration algorithm.

As can be seen, passing the  gradient through the power iteration involves rather elementary operations, such as the derivative of a matrix inverse. For those complex-valued gradients, the mapping to real operations works just fine.



\section{Training Objectives}
\label{sec:training_objectives}






In the following description of loss functions we 
ignore the source permutation problem in the notation for better readability.

\subsection{Frequency-Dependent SDR}
As a reference, we first consider a loss function that is defined in the \gls{STFT} domain, the mean squared error between the complex-valued target signal, $d_{t, f, i, r}$, and its estimate obtained at the beamformer output $\hat{d}_{t, f, i, r}$:

\begin{align}
    \label{eq:L_F-SDR}
    L^{\mathrm{F-SDR}}
    &= \frac{10}{I}\sum_{i} \log_{10}\left(\frac{\sum_{t, f} | d_{t, f, i, r} - \hat{d}_{t, f, i, r}|^2}{\sum_{t, f} |d_{t, f, i, r} |^2}\right) ,
\end{align}
where the denominator, which is a constant w.r.t. network parameters, is  introduced  for better interpretability of the loss as a frequency-dependent signal-to-distortion ratio\footnote{Since \glspl{NN} want to minimize the loss function, the loss function is the negative \gls{SDR}.}.
Note that this loss is equivalent to the phase sensitive loss \cite{Kolbaek2017uPIT}, except for the log operation and a scale.
\subsection{SDR}
Casting
the loss of \cref{eq:L_F-SDR} to the time-domain, we obtain the time-domain \gls{SDR} loss,
\begin{align}
    L^{\mathrm{SDR}}
    &= \frac{10}{I}\sum_{i} \log_{10}\left(\frac{\sum_{\ell} | d_{\ell, i, r} - \hat{d}_{\ell, i, r}|^2}{\sum_{\ell} |d_{\ell, i, r} |^2}\right),
\end{align}
which is the loss proposed in \cite{Heitkaemper2020demystifying}.
\subsection{Scale Invariant SDR}
In \cite{Heitkaemper2020demystifying} it is also shown that this loss is closely related to the \gls{SISDR} loss used in
 \cite{Luo2018tasnet}:
\begin{align}
    \label{eq:t-si-sdr}
    L^{\mathrm{SI-SDR}}
    &= \frac{10}{I}\sum_{i} \log_{10}\left(\frac{\sum_{\ell} | d_{\ell, i, r} \hat{a}_i - \hat{d}_{\ell, i, r}|^2}{\sum_{\ell} |d_{\ell, i, r} \hat{a}_i |^2}\right) ,
\end{align}
where the scaling term $\hat{a}_i$ is introduced to compensate for a potential scaling error between target and estimate. It is computed as
\begin{align}
    \hat{a}_i &= \argmin_{a_i} \left\{\sum_{\ell}\left| d_{\ell, i, r} a_i - \hat{d}_{\ell, i,r} \right|^2 \right\}.
\end{align}
\subsection{Convolutive transfer function Invariant SDR}
Separation networks trained with the \gls{SISDR} loss have been reported to deliver very good separation results \cite{Luo2018tasnet,Kolbaek2020loss}.
However, those observations have been mostly made in single-channel scenarios.
When considering a multi-channel reverberant setup, it was shown in \cite{Drude2019sms} that \gls{SISDR} produces strange artifacts.
For example, if the \gls{SISDR} is calculated between one channel as estimate and another channel as target, i.e., they have a different \gls{RIR}, there is no audible difference, but the \gls{SISDR} indicates a huge difference \cite{Drude2019sms}.

This observation led to the opinion that the invariance to a short impulse response that is given in the original \gls{SDR} measure of the BSS Eval toolbox \cite{Vincent2006BSSEval}, and that was criticized in \cite{LeRoux2019SDR}, is actually beneficial in reverberant scenarios.
We therefore propose the following  training objective, which we call  \gls{CISDR}:
\begin{align}
    \label{eq:t-ci-sdr}
    L^{\mathrm{CI-SDR}} &= \frac{10}{I}\sum_{i} \log_{10}\left(\frac{\sum_{\ell} | \sum_{\tau} s_{\ell-\tau, i}  \hat{a}_{\tau, i} - \hat{d}_{\ell, i}|^2}{\sum_{\ell} |\sum_{\tau}s_{\ell - \tau, i} \hat{a}_{\tau, i} |^2}\right) , \\
    \hat{a}_{\tau,i} &= \argmin_{a_{\tau, i}} \left\{\sum_{\ell}\left| \sum_{\tau} s_{\ell - \tau, i} a_{\tau, i} - \hat{d}_{\ell, i} \right|^2 \right\} .
\end{align}
where the estimation of $\hat{a}_{\tau, i}$ uses a solution of the Wiener-Hopf equation.
Note that, unlike in \gls{SISDR}, $\hat{a}_{\tau, i}$ is a finite impulse response filter with 512 coefficients\footnote{To be precise, we reimplemented "BSS Eval v3" with gradient support.}.
As far as we know, this is the first time, that \gls{CISDR} is used as a training objective.
Both \cite{Drude2019sms,LeRoux2019SDR} tried to interpret them as metric, 
here we want to compare them as \gls{NN} training objectives.

In \cite{LeRoux2019SDR}, the authors argued that the \gls{CISDR} measure of the BSS Eval toolbox does not punish all kinds of distortions, e.g., the suppression of some frequencies.
Instead, they proposed to use \gls{SISDR} in scenarios of anechoic single channel mixture recordings.
Nevertheless, the comparison of \gls{SISDR} with \gls{CISDR} in more realistic scenarios without reverberation did not show a clear advantage of one metric over the other \cite{LeRoux2019SDR}.

Note that we here consider a different scenario, multi-channel reverberant recordings.
We argue that the above issue of completely suppressing some frequencies cannot occur because of the regularizing effect of the MVDR beamformer.
Its distortionless constraint ensures that unwanted solutions that may still drive the \gls{SDR} to large values are not allowed.   



Another difference between eq.~\eqref{eq:t-ci-sdr} and \eqref{eq:t-si-sdr}:
\gls{SISDR} needs a target signal that is aligned with the input. So, a reverberated signal is used as target.
Here, we follow \cite{Heymann2017generic}, who used just the early part (i.e., \SI{50}{ms}) of the \gls{RIR} to generate the target.
In eq.~\eqref{eq:t-ci-sdr} we use the source signal $s_{\ell, i}$ convolved with the best (in the \gls{MSE} sense) \gls{RIR}, of the length \SI{32}{ms}, as target.
We conjecture that this drives the beamformer to have also a dereverberating effect. 

\section{Experiments}
\label{sec:experiments}

\subsection{Dataset}
For the experiments we used a simulated dataset that was developed during the JSALT 2020 workshop at JHU.
This dataset uses the clean utterances from LibriSpeech \cite{Panayotov2015librispeech} at a sample rate of \SI{16}{kHz} and generates \SI{600}{\hour}, \SI{3.6}{\hour} and \SI{3.6}{\hour} of training, development and testing data, respectively.
Each mixture contains 2 utterances which either have full or partial overlap.
The reverberation time $T_{60}$ and \gls{SNR} (spherical noise) is uniformly sampled from \SI{0.15}{\second} to \SI{0.6}{\second} and \SI{10}{\decibel} to \SI{20}{\decibel}, respectively.
The \glspl{RIR} are generated by image method \cite{Allen1979imageMethode,Habets2006RIRgenerator} for a circular array with $7$ microphones and minimum angle between the speakers is \SI{5}{\degree}.

\begin{table}[t]
    \centering
    \caption{
        Scores for no enhancement, beamforming with oracle masks (Wiener Like Mask (WLM) \cite{Erdogan2015phase}) and Eq.~\eqref{eq:rtfest},  and directly using oracle signals as ASR input.
    }\label{tab:separationReference}
    \setlength{\tabcolsep}{2.8pt}
    \renewcommand*{\arraystretch}{1.05}
    
    \begin{tabular}{
            l 
            l 
            l
            S[round-precision=2,round-mode=places, table-format = 1.2]
            S[round-precision=2,round-mode=places, table-format = 2.2]
            S[round-precision=3,round-mode=figures, table-format = 1.3]
            S[round-precision=1,round-mode=places, table-format = 2.1]
            }
        \toprule
        {Signal} & {Mask} & {Enh.}
        & {PESQ} & {BSS Eval} & {STOI} & {WER} \\
        & &
        & & {SDR} & & \\
        \midrule
        Obs. $y_{\ell, r}$ & {---} & {---} & 1.2185274861335755 & -0.475615083140195 & 0.7152954227067454 & 96.400000 \\ 
        \midrule
        Obs. $\y_{t, f}$ & WLM & MVDR(eig) & 2.256359 & 16.372279 & 0.910595 & 3.800000 \\ 
        \midrule
        Early $\d_{\ell, i}$ & {---} & {---} & 2.7632231127262115 & 19.255025346063107 & 0.919621085254769 & 3.200000 \\ 
        Source $s_{\ell, i}$ & {---} & {---} & 4.643888473510742 & 289.47550274611814 & 0.9999999999999997 & 3.000000 \\ 
        \bottomrule
    \end{tabular}
 \end{table}
 
\subsection{Results}
\glsreset{WER}
For judging the trained systems, we compare 4 metrics: 
\gls{PESQ} \cite{Rix2001PESQ}, BSS Eval SDR \cite{Vincent2006BSSEval}, \gls{STOI} \cite{Taal2011STOI} and \gls{WER}.
For the speech enhancement metrics, \gls{PESQ}, BSS Eval SDR and \gls{STOI}, the source signal $s_{\ell, i}$ is used as reference.
The BSS Eval SDR has to be taken with care, because the \gls{CISDR} objective tries to optimize this score.
For the \gls{WER} calculation we used a pretrained system from ESPnet \cite{Watanabe2018espnet}, a transformer-based \gls{ASR} system \cite{Karita2019ASR}.
Note that the \gls{ASR} system was only trained on clean utterances and was not adapted to any enhancement artifacts.
\Cref{tab:separationReference} shows results of some  reference systems. Without any enhancement the \gls{WER} is close to \SI{100}{\percent}.
Using oracle masks and beamforming, the \gls{WER} can be drastically reduced.
The final two rows show the performance for the source signal $s_{\ell, i}$ and the early signal $d_{\ell, i, r}$ as input to the \gls{ASR} system, respectively.

In \cref{tab:separationBeamformers} we compare the performance of different loss functions.
As a reference, the first row displays the performance of a classical \gls{PIT} system.
The mask estimator is trained with a frequency loss, \cref{eq:L_F-SDR}, and source extraction is done by masking.
In the second row the enhancement at test time is changed to a beamformer.

The other systems in this table use a beamformer for source extraction, both at training and test time.
It can be observed that the system trained with the  \gls{CISDR} criterion clearly outperforms the other systems.

If the three power iterations for dominant eigenvector estimation on the test data are replaced by an eigenvector decomposition the \gls{WER} can be further slightly reduced to $\SI{4.2}{\%}$.
This is only \num{1.2} percentage points worse than the \gls{WER} on single-source non-reverberant input, shown in \cref{tab:separationReference}.
Comparing this system with an oracle mask (second row of Table~\ref{tab:separationReference}), our system outperforms it in terms of the speech enhancement metrics and comes close in terms of \gls{WER} performance: the WER of the oracle mask-based system is only \num{0.4} percentage points better.

\begin{table}[t]
    \centering
    \caption{
        Separation performance of enhancement (Enh.) systems that use masking or \gls{MVDR} beamforming.
        In \textit{MVDR} the \gls{RTF} estimation is done with $3$ power iterations, while \textit{MVDR(eig)} uses the eigenvalue decomposition.
        The training loss is varied from frequency \gls{SDR} over time-domain \gls{SDR} to Convolutive transfer function Invariant \gls{SDR}.
    }\label{tab:separationBeamformers}
    \setlength{\tabcolsep}{2.8pt}
    \renewcommand*{\arraystretch}{1.05}
    
    \begin{tabular}{
            l
            l
            l
            S[round-precision=2,round-mode=places, table-format = 1.2]
            S[round-precision=2,round-mode=places, table-format = 2.2]
            S[round-precision=3,round-mode=figures, table-format = 1.3]
            S[round-precision=1,round-mode=places, table-format = 2.1]
        }
        \toprule
        \multicolumn{2}{l}{Train} & Test & {PESQ} & {BSS Eval} & {STOI} & {WER} \\
        \cmidrule(l{2pt}r{2pt}){1-2}
        \cmidrule(l{2pt}r{2pt}){3-3}
        Enh.  & Loss & Enh. & & {SDR} & & \\
        \midrule
        Masking & F-SDR & Masking & 1.488299 & 8.371823 & 0.786646 & 45.600000 \\
        Masking & F-SDR & MVDR(eig) & 1.938309 & 13.633940 & 0.884185 & 8.200000 \\
        \midrule
        MVDR & F-SDR & MVDR & 1.985229 & 15.375089 & 0.892858 & 7.900000 \\
        MVDR & SDR & MVDR & 1.976216 & 15.076299 & 0.892770 & 6.700000 \\
        MVDR & SI-SDR & MVDR & 2.011745 & 15.575882 & 0.894783 & 6.900000 \\
        MVDR & CI-SDR & MVDR & 2.464840 & 20.404034 & 0.929792 & 4.400000 \\
        \midrule
        MVDR & CI-SDR & MVDR(eig) & 2.50 & 20.61 & 0.93 & 4.2 \\
        \bottomrule
    \end{tabular}
 \end{table}

\section{Conclusions}
\label{sec:conclusions}

\glsreset{CISDR}
This paper proposes to use a \gls{CISDR}, i.e., BSS Eval \gls{SDR}, as the training criterion of a \gls{NN} supported multi-channel beamforming based source separation system.
The effectiveness is shown on an artificially mixed reverberant speech database.
It outperforms a classical \gls{PIT} system, irrespective of whether source extraction is done by masking or by beamforming at test time in that system, but also outperforms systems trained on alternative time-domain objectives.
Furthermore, the system is compared with an oracle mask definition, where it outperforms the oracle mask in the speech enhancement metrics and approaches it in terms of \gls{WER} performance.
The final \gls{WER} is \SI{4.2}{\percent}, while the \gls{WER} of \SI{3.0}{\percent} on the non-reverberant single speaker source signals is only \SI{1.2}{\percent} better.
To achieve this no fine tuning of the \gls{ASR} system was necessary.


\section{Acknowledgements}
We deeply thank Prof. Shinji Watanabe for many helpful discussions.
Computational resources were provided by the Paderborn Center for Parallel Computing.
The work reported here was started at JSALT 2020 at JHU, with support from Microsoft, Amazon and Google.
\pagebreak
\let\OLDthebibliography\thebibliography
\renewcommand\thebibliography[1]{
  \OLDthebibliography{#1}
  \setlength{\itemsep}{0.5pt plus 0.3ex}
}
\bibliographystyle{IEEEbib}
\bibliography{strings,refs}

\begin{thebibliography}{10}

\bibitem{Comon1994ICA}
P.~Comon,
\newblock ``Independent component analysis, a new concept?,''
\newblock {\em Signal processing}, vol. 36, no. 3, pp. 287--314, 1994.

\bibitem{Kim2006IVA}
T.~Kim, H.~T. Attias, S.-Y. Lee, and T.-W. Lee,
\newblock ``Blind source separation exploiting higher-order frequency
  dependencies,''
\newblock {\em IEEE Transactions on Audio, Speech, and Language Processing},
  vol. 15, no. 1, pp. 70--79, 2006.

\bibitem{Lee1999NNMF}
D.~D. Lee and H.~S. Seung,
\newblock ``Learning the parts of objects by non-negative matrix
  factorization,''
\newblock {\em Nature}, vol. 401, no. 6755, pp. 788--791, 1999.

\bibitem{Araki2006SMM}
S.~Araki, H.~Sawada, R.~Mukai, and S.~Makino,
\newblock ``Normalized observation vector clustering approach for sparse source
  separation,''
\newblock in {\em 2006 14th European Signal Processing Conference}. IEEE, 2006,
  pp. 1--5.

\bibitem{Yilmaz2004WDO}
O.~Yilmaz and S.~Rickard,
\newblock ``Blind separation of speech mixtures via time-frequency masking,''
\newblock {\em IEEE Transactions on signal processing}, vol. 52, no. 7, pp.
  1830--1847, 2004.

\bibitem{Pertila2015distant}
P.~Pertil{\"a} and J.~Nikunen,
\newblock ``Distant speech separation using predicted time--frequency masks
  from spatial features,''
\newblock {\em Speech communication}, vol. 68, pp. 97--106, 2015.

\bibitem{Drude2019integration}
L.~Drude and R.~Haeb-Umbach,
\newblock ``Integration of neural networks and probabilistic spatial models for
  acoustic blind source separation,''
\newblock {\em IEEE Journal of Selected Topics in Signal Processing}, vol. 13,
  no. 4, pp. 815--826, 2019.

\bibitem{Kolbaek2017uPIT}
M.~Kolb{\ae}k, D.~Yu, Z.-H. Tan, and J.~Jensen,
\newblock ``Multitalker speech separation with utterance-level permutation
  invariant training of deep recurrent neural networks,''
\newblock {\em IEEE/ACM Transactions on Audio, Speech, and Language
  Processing}, vol. 25, no. 10, pp. 1901--1913, 2017.

\bibitem{Hershey2016deepClustering}
J.~R. Hershey, Z.~Chen, J.~Le~Roux, and S.~Watanabe,
\newblock ``Deep clustering: Discriminative embeddings for segmentation and
  separation,''
\newblock in {\em ICASSP}. IEEE, 2016, pp. 31--35.

\bibitem{Wang2018}
Z.~{Wang}, J.~{Le Roux}, and J.~R. {Hershey},
\newblock ``Multi-channel deep clustering: Discriminative spectral and spatial
  embeddings for speaker-independent speech separation,''
\newblock in {\em ICASSP}. IEEE, 2018, pp. 1--5.

\bibitem{Luo2018tasnet}
Y.~Luo and N.~Mesgarani,
\newblock ``{TasNet}: Time-domain audio separation network for real-time,
  single-channel speech separation,''
\newblock in {\em ICASSP}. IEEE, 2018, pp. 696--700.

\bibitem{Kolbaek2020loss}
M.~Kolb{\ae}k, Z.-H. Tan, S.~H. Jensen, and J.~Jensen,
\newblock ``On loss functions for supervised monaural time-domain speech
  enhancement,''
\newblock {\em IEEE/ACM Transactions on Audio, Speech, and Language
  Processing}, vol. 28, pp. 825--838, 2020.

\bibitem{Heitkaemper2020demystifying}
J.~Heitkaemper, D.~Jakobeit, C.~Boeddeker, L.~Drude, and R.~Haeb-Umbach,
\newblock ``Demystifying {TasNet}: A dissecting approach,''
\newblock in {\em ICASSP}. IEEE, 2020, pp. 6359--6363.

\bibitem{Chang2019mimo}
X.~Chang, W.~Zhang, Y.~Qian, J.~Le~Roux, and S.~Watanabe,
\newblock ``{MIMO-Speech}: End-to-end multi-channel multi-speaker speech
  recognition,''
\newblock in {\em ASRU}. IEEE, 2019, pp. 237--244.

\bibitem{Heymann2017beamnet}
J.~Heymann, L.~Drude, C.~Boeddeker, P.~Hanebrink, and R.~Haeb-Umbach,
\newblock ``Beamnet: End-to-end training of a beamformer-supported
  multi-channel {ASR} system,''
\newblock in {\em ICASSP}. IEEE, 2017, pp. 5325--5329.

\bibitem{Heymann2018MBBFSmartHome}
J.~Heymann, M.~Bacchiani, and T.~N. Sainath,
\newblock ``Performance of mask based statistical beamforming in a smart home
  scenario,''
\newblock in {\em ICASSP}. IEEE, 2018, pp. 6722--6726.

\bibitem{Drude2019sms}
L.~Drude, J.~Heitkaemper, C.~Boeddeker, and R.~Haeb-Umbach,
\newblock ``{SMS-WSJ}: Database, performance measures, and baseline recipe for
  multi-channel source separation and recognition,''
\newblock {\em arXiv preprint arXiv:1910.13934}, 2019.

\bibitem{Vincent2006BSSEval}
E.~Vincent, R.~Gribonval, and C.~F{\'e}votte,
\newblock ``Performance measurement in blind audio source separation,''
\newblock {\em IEEE Transactions on Audio, Speech, and Language Processing},
  vol. 14, no. 4, pp. 1462--1469, 2006.

\bibitem{Panayotov2015librispeech}
V.~Panayotov, G.~Chen, D.~Povey, and S.~Khudanpur,
\newblock ``Librispeech: An {ASR} corpus based on public domain audio books,''
\newblock in {\em ICASSP}. IEEE, 2015, pp. 5206--5210.

\bibitem{Ban1988beamforming}
B.~D. Van~Veen and K.~M. Buckley,
\newblock ``Beamforming: A versatile approach to spatial filtering,''
\newblock {\em IEEE ASSP Magazine}, vol. 5, no. 2, pp. 4--24, 1988.

\bibitem{Ito2017MaxEigATF}
N.~Ito, S.~Araki, M.~Delcroix, and T.~Nakatani,
\newblock ``Probabilistic spatial dictionary based online adaptive beamforming
  for meeting recognition in noisy and reverberant environments,''
\newblock in {\em ICASSP}. IEEE, 2017, pp. 681--685.

\bibitem{Boeddeker2017NNBF}
C.~Boeddeker, P.~Hanebrink, L.~Drude, J.~Heymann, and R.~Haeb-Umbach,
\newblock ``Optimizing neural-network supported acoustic beamforming by
  algorithmic differentiation,''
\newblock in {\em ICASSP}. IEEE, 2017, pp. 171--175.

\bibitem{Mises1929PowerIteration}
R.~Mises and H.~Pollaczek-Geiringer,
\newblock ``Praktische {Verfahren} der {Gleichungsaufl{\"o}sung}.,''
\newblock {\em ZAMM-Journal of Applied Mathematics and Mechanics/Zeitschrift
  f{\"u}r Angewandte Mathematik und Mechanik}, vol. 9, no. 1, pp. 58--77, 1929.

\bibitem{LeRoux2019SDR}
J.~Le~Roux, S.~Wisdom, H.~Erdogan, and J.~R. Hershey,
\newblock ``{SDR}--half-baked or well done?,''
\newblock in {\em ICASSP}. IEEE, 2019, pp. 626--630.

\bibitem{Heymann2017generic}
J.~Heymann, L.~Drude, and R.~Haeb-Umbach,
\newblock ``A generic neural acoustic beamforming architecture for robust
  multi-channel speech processing,''
\newblock {\em Computer Speech \& Language}, vol. 46, pp. 374--385, 2017.

\bibitem{Allen1979imageMethode}
J.~B. Allen and D.~A. Berkley,
\newblock ``Image method for efficiently simulating small-room acoustics,''
\newblock {\em The Journal of the Acoustical Society of America}, vol. 65, no.
  4, pp. 943--950, 1979.

\bibitem{Habets2006RIRgenerator}
E.~A. Habets,
\newblock ``Room impulse response generator,''
\newblock {\em Technische Universiteit Eindhoven, Tech. Rep}, vol. 2, no. 2.4,
  pp. 1, 2006.

\bibitem{Erdogan2015phase}
H.~Erdogan, J.~R. Hershey, S.~Watanabe, and J.~Le~Roux,
\newblock ``Phase-sensitive and recognition-boosted speech separation using
  deep recurrent neural networks,''
\newblock in {\em ICASSP}. IEEE, 2015, pp. 708--712.

\bibitem{Rix2001PESQ}
A.~W. Rix, J.~G. Beerends, M.~P. Hollier, and A.~P. Hekstra,
\newblock ``Perceptual evaluation of speech quality ({PESQ})-a new method for
  speech quality assessment of telephone networks and codecs,''
\newblock in {\em ICASSP}. IEEE, 2001, vol.~2, pp. 749--752.

\bibitem{Taal2011STOI}
C.~H. Taal, R.~C. Hendriks, R.~Heusdens, and J.~Jensen,
\newblock ``An algorithm for intelligibility prediction of time--frequency
  weighted noisy speech,''
\newblock {\em IEEE Transactions on Audio, Speech, and Language Processing},
  vol. 19, no. 7, pp. 2125--2136, 2011.

\bibitem{Watanabe2018espnet}
S.~Watanabe, T.~Hori, S.~Karita, T.~Hayashi, J.~Nishitoba, Y.~Unno, N.~{Enrique
  Yalta Soplin}, J.~Heymann, M.~Wiesner, N.~Chen, A.~Renduchintala, and
  T.~Ochiai,
\newblock ``{ESPnet}: End-to-end speech processing toolkit,''
\newblock in {\em Proceedings of Interspeech}, 2018, pp. 2207--2211.

\bibitem{Karita2019ASR}
S.~Karita, N.~Chen, T.~Hayashi, T.~Hori, H.~Inaguma, Z.~Jiang, M.~Someki,
  N.~E.~Y. Soplin, R.~Yamamoto, X.~Wang, et~al.,
\newblock ``A comparative study on transformer vs {RNN} in speech
  applications,''
\newblock in {\em ASRU}. IEEE, 2019, pp. 449--456.

\end{thebibliography}
\balance

\end{document}